# Forecasting Inflation Based on Hybrid Integration of the Riemann Zeta Function and the FPAS Model (FPAS + ζ): Cyclical Flexibility, Socio-Economic Challenges and Shocks, and Comparative Analysis of Models

**Davit Gondauri,** **ORCID:** https://orcid.org/0000-0002-9611-3688
PhD, Professor, Doctor of Business Administration, Business & Technology University, Georgia

**Corresponding author:** Davit Gondauri, dgondauri@gmail.com
**Type of manuscript:** research paper

**Abstract:** *Inflation forecasting is one of the main socio-economic challenges in modern macroeconomic modeling, especially when cyclical, structural, and shock factors act simultaneously in the system. Traditional forecasting systems, such as Forecasting and Policy Analysis System (FPAS) and Autoregressive Integrated Moving Average (ARIMA), often fail to provide an adequate analysis of cyclical asymmetry and unexpected fluctuations. The presented study proposes a hybrid approach that combines a structural macroeconomic model (FPAS) and the cyclical components of the Riemann zeta function, creating a completely new forecasting framework (FPAS + ζ). The study is the first to use the cyclical components of the Riemann zeta function in practice within a structural macroeconomic model, which significantly increases the accuracy of the forecast and the ability to adapt to policy instruments. The uniqueness of the model lies in the fact that it reduces the forecast error and increases the system's responsiveness to cyclical and shock conditions. The cyclical characteristic is modeled based on the macroeconomic indicators of Georgia (2005–2024), using a nonlinear argument t, on which the cyclical adjustment is based on the formula ζ(0.5 + i·t). The forecast is modulated by an α-coefficient, which is optimized by the principle of Root Mean Square Error (RMSE) minimization. The model also combines Fourier decomposition for spectral analysis and the hidden Markov model for phase identification, which provides a deep analysis of inflation fluctuations over time. The implementation of this hybrid inflation forecasting framework will make a significant contribution to solving socio-economic challenges, thereby increasing the accuracy of forecasting and flexibility at the practical level of policy.*

**Received:** 21.07.2025      **Accepted:** 22.09.2025      **Published:** 03.10.2025

**Funding:** There is no funding for this research.
**Publisher:** Academic Research and Publishing UG (i.G.) (Germany).
**Founder:** Academic Research and Publishing UG (i.G.) (Germany).

**Cite as:** Gondauri, D. (2025). Forecasting Inflation Based on Hybrid Integration of the Riemann Zeta Function and the FPAS Model (FPAS + ζ): Cyclical Flexibility, Socio-Economic Challenges and Shocks, and Comparative Analysis of Models. *SocioEconomic Challenges*, *9*(3), 1-20. https://doi.org/10.61093/sec.9(3).1-20.2025

**INTRODUCTION**

Inflation forecasting is one of the most complex and strategic socio-economic challenges in modern macroeconomics. Its complexity stems not only from the multifactorial nature of inflation but also from the fact that inflation combines structural, cyclical, and shock components, the interaction of which often does not lend itself to linear and static assessments. As a result, forecasting efforts based on historical data and linear trends often fail to reflect the dynamic nature of the economy and fail to provide an adequate response to external or internal shocks.

It is crucial for economic policy-making institutions – especially the central bank and the ministry of finance – to have an adaptable and multi-layered forecasting framework in order to effectively respond to ongoing socio-economic challenges, as well as to identify a range of viable scenarios.

In this direction, widely used forecasting models, such as the Forecasting and Policy Analysis System (FPAS) and ARIMA, are based primarily on trends and linear relationships built on past data. However, such approaches are limited in their predictive flexibility – especially during periods of crisis, such as the global financial crisis (2009), the sharp devaluation of the lari (2015), or the economic shutdown caused by the COVID-19 pandemic (2020).

It is precisely at this juncture that a radical, theoretically strengthened, and interdisciplinary approach emerges: the integration of the Riemann zeta function into the inflation forecasting framework. The Riemann zeta function ($\zeta(s)$) is one of the deepest and most complex objects in mathematics; the arrangement of its zeros on the critical line $Re(s) = 0.5$ is known as the Riemann hypothesis. The hypothesis is a grandiose mathematical problem, included in the list of Millennium Problems of the Clay Mathematics Institute and has been exciting theoretical science for two centuries.

Despite its deep abstraction, this function has periodicity, cyclicality, and spectral density properties that correlate with the nature of economic cycles. In particular, the distribution of zeros on the critical line creates a possible framework for the mathematical treatment of economic phases, fluctuations, and resonances – which has so far been carried out mainly at the empirical or hypothetical level.

The present study proposes a hybrid framework for inflation forecasting that combines the structural macromodel FPAS and the cyclical components of the Riemann zeta function. This model allows not only for adding a cyclical correction to the forecast but also for creating an adaptive mechanism for automatic adjustment of the forecast according to economic signals. Practically, this is achieved by forming a nonlinear argument $t$ based on Georgia's macroeconomic parameters (2005–2024): real GDP, money supply, monetary rate, exchange rate, unemployment and other indicators. The above $t$ is introduced into the expression $\zeta(0.5 + i\,t)$ and generates a cyclical correction of the inflation forecast. This approach is not only an extension of the theoretical framework for forecasting, but also an example of the convergence of scientific knowledge, where higher mathematics is transferred to applied economic analysis. At a time when economic science is constantly searching for new methods for assessing global instability and policy effectiveness, the interpretation of the Riemann hypothesis in economic modeling can be considered an innovative scientific transformation – an intellectual bridge between mathematics and economic reality. Overall, the presented model aims to address current and future socio-economic challenges facing economic policy-makers.

**Research Hypothesis, Objectives and Tasks**

*Research Hypothesis*

$H_0$ (Null Hypothesis): The FPAS + $\zeta$ model, which combines the cyclical component of the Riemann zeta function and a traditional macroeconomic model (FPAS), does not provide a better forecast of inflation dynamics than the separate FPAS or ARIMA models.

$H_1$ (Alternative Hypothesis): The FPAS + $\zeta$ model demonstrates higher forecast accuracy, sensitivity to cyclicality and adaptability in responding to economic shocks than traditional models (FPAS, ARIMA).

*Research Objective*

The objective is to develop, validate and evaluate a hybrid macroeconomic model (FPAS + $\zeta$) that uses the cyclical component of the Riemann zeta function to forecast inflation and to compare its effectiveness with traditional forecasting models.

*Research Tasks*

- Model framework formation: creation of the theoretical architecture of the hybrid model (FPAS + ζ) and integration of the zeta function.
- Transformation of economic variables: conversion of macroeconomic parameters into *t*-variables and determination of their functional relationship.
- Model calibration: optimization of the modulation coefficient α by minimizing the RMSE.
- Cyclicality and shock analysis: estimation of the cyclical flexibility and response of the model using Fourier analysis.
- Identification of phase structure: use of hidden Markov model to predict inflation phases.
- Integration of stochastic shocks: addition of stochastic component of *ε(t)*, and conducting Monte Carlo simulation.
- Comparative analysis of models: estimation of FPAS, FPAS + ζ and ARIMA models using the Analytic Hierarchy Process (AHP)-Technique for Order Preference by Similarity to Ideal Solution (TOPSIS) method.
- Determining the potential for practical application: proposing a scheme for using the model for policy-making institutions.

The uniqueness of this research is expressed in the integration of the Riemann ζ function into the inflation forecasting framework, which offers a completely new, interdisciplinary vision to the academic community and policy-making institutions compared to the traditional approaches that have existed so far. In particular, the FPAS + ζ model creates a new horizon for the interaction of mathematics and economics and allows for high-accuracy forecasting of cyclicality, which is important in modern macroeconomic modeling.

**LITERATURE REVIEW**

Inflation forecasting is one of the central tasks of modern macroeconomics, which requires constant refinement of models taking into account cyclical, structural, and shock factors. Along with traditional forecasting frameworks, there is a growing interest in the use of mathematical functions – especially the Riemann zeta function – in economic modeling, which reveals the cyclical depth and complexity of forecasts. Addressing socio-economic challenges remains a key motivation for the ongoing development of these advanced forecasting methodologies. Spectral analysis and Fourier decomposition of macroeconomic series enhance the perception of cyclicality, while the use of hidden Markov model expands the ability to identify phases and respond to systemic changes. In addition, AHP and TOPSIS methods are useful for comparative evaluation of models and optimization of economic decisions. As a result, studies combine macromodeling, mathematical analysis, and multi-criteria estimation, creating an interdisciplinary framework for inflation forecasting and policy analysis.

*Inflation Forecasting Frameworks and Model Structure*

Inflation forecasting remains a central challenge in macroeconomic policy. Laxton et al. (2009) emphasize the early introduction of forecasting frameworks and their institutional improvement over time. IMF reports (2021, 2024) focus on both model-based policymaking and frameworks adapted to the needs of developing countries. Al-Mashat (2018) assesses the effectiveness of operational components, especially in environments where the management of inflation expectations is critical. This process is developed by Ahmad et al. (2017), which combines fiscal and external blocs in a small-scale DSGE model, while the National Bank of Kazakhstan (2024) creates a mixed forecasting model that combines elements of a traditional econometric framework and machine learning. Alternative forecasting mechanisms are being worked on by Shibata (2019), who uses a hidden Markov model to describe cyclical transitions in the labor market and shows its superiority over the FOM model. In terms of assessing the accuracy of forecasts, Malik and Hanif (2018) compare multivariate models, where ARDL and trimmed mean gain advantages depending on the inflation regime. Pandey et al. (2021) use text analysis to illustrate systematic deviations in Indian forecasts, which confirms the governance need for structural reform. The same topic is developed by Nikolaychuk and Sholomytskyi (2015), who discuss the modeling practices of the National Bank of Ukraine and the need to strengthen the forecasting system. These studies highlight the importance of model innovation in response to evolving socio-economic challenges that impact inflation dynamics and policy effectiveness.

*The Riemann Zeta Function in Economic Modeling*

The potential of the Riemann Zeta Function is increasingly apparent in the search for new ways of economic modeling. Chaudhry et al. (2011) describe the local structure of zero points with a generalized version of the function. Thalassinakis (2025) relates the zeta function to a new theory of infinities, thereby proving the correlation properties of zeros, while Belovas (2023) discusses the characteristics of the absolute values of the function at symmetric points. Borwein et al. (2000) provide a detailed overview of technical strategies for calculating the zeta, including convergent sequences and value recycling techniques. Bettin (2010) develops an asymptotic analysis of the second moment of the function at high displacements, which creates a new basis for modeling its dynamics. Ewerhart (2024) builds game theory models based on the zeta function, the stability of which depends on the truth of the hypothesis. An attempt to integrate multiple theoretical paradigms is found in the work of Mahmoud (2025), who combines spectral, probabilistic, and analytical approaches into a single framework. Da Silva (2025) connects the zeros of zeta with the differences of primes, on the basis of which he builds a model of causal networks. A practical economic interpretation is offered by Caporali (2023), who considers the phase cyclicality of the market based on the distances between the zeros of zeta and demonstrates the applicability of the GUE-distribution in financial forecasting.

*Spectral Analysis and Fourier Decomposition in Macroeconomics*

Zhang (2016) focuses on spectral analysis of high-frequency financial series and argues its superiority over traditional models. Bezuidenhout et al. (2021) investigates the cyclicality of value investing, which is a reflection of the adaptive market hypothesis. Li et al. (2023) describe structural disruptions in financial markets through adaptive Fourier decomposition (AFD). Jun et al. (2019) focus on the early detection of crisis phases through frequency analysis of Fourier modules. The performance of the model on mass series is demonstrated by Masset (2008), who analyzes both Fourier and wavelet transforms. Conway and Frame (2000) compare different filtering methods in the cyclical analysis of output gaps. Pollock (2008) discusses the matrix formalism of Fourier theory, while Cochrane (1997/2005) proposes its practical application with a focus on the scalar and Wold theorem. The integration of historical context is suggested by Granger and Watson (1984), who analyze the time and spectral methods as two complementary paradigms.

*Hidden Markov model in Economic Phases*

Du et al. (2020) construct an earnings fidelity metric using a Hidden Markov Model (HMM) that describes the correspondence of a company's financial signal to the real situation. Nguyen (2018) uses an HMM to evaluate a trading strategy where the forecast exceeds the historical average. Kim et al. (2019) model the dynamics of global assets with an HMM structure and capture cyclical market reactions. Bucci et al. (2021) combine R&D-based growth theory with HMM regimes to provide a phased assessment of innovation in OECD countries. Yuan and Mitra (2016) use HMM to identify bull/bear regimes, where the model captures volatility clustering better than the classical GBM framework.

*Comparative evaluation of models: AHP and TOPSIS*

On the issue of multi-criteria decision support, Ccatamayo-Barrios et al. (2023) compare AHP and TOPSIS in selecting the optimal mining method. Rahman (2024) develops the effectiveness of the models in retail location selection systems. Vincent (2023) shows that under similar criteria, AHP and TOPSIS give identical results in the context of bearing puller design. Zhu et al. (2019) implement a hybrid integration of these methods for the evaluation of urban renewal projects, resulting in a sustainable development support system through a complex systematization of criteria prioritization and ranking.

This literature review shows that inflation forecasting frameworks are constantly evolving through the integration of both structural and mathematically complex models. Along with traditional macroeconomic frameworks, there is growing interest in complex mathematical tools such as the Riemann zeta function, spectral analysis, hidden Markov models, and multi-criteria decision-making methodologies. Bridging the gap between the theoretical heights of the zeta function and the macroeconomic interpretation of the behavior of phase cycles is an important direction, especially when the subject of forecasting is the phase changes of inflation dynamics and shock identification.

On this basis, the present study develops a hybrid forecasting model that combines the traditional FPAS framework with the cyclical components (ζ) of the Riemann zeta function. The research strategy is based on a combination of three main stages:
• defining a structural model within the framework of FPAS,
• extracting cyclical signals based on ζ(0.5 + i t) using spectral and Fourier analysis,
• comparative evaluation of models using HMM, AHP and TOPSIS methods in terms of shock detection, cyclicality forecasting and phase change identification.

As a result, an integrated forecasting framework (FPAS + ζ) emerges, which improves the modeling of inflationary processes by taking into account cyclical flexibility, dynamic adaptation, and nonlinear components. This strategy develops both theoretical and applied perspectives for economic policy planning, improving the mechanism of response to shocks and increasing the long-term accuracy of forecasts. Such frameworks are critical for anticipating and managing socio-economic challenges in both national and global contexts.

**METHODOLOGY**

*Research Type and Methodological Approach*

The presented study is a mixed method, combining quantitative and qualitative approaches. The quantitative analysis is based on modeling the cyclical component of the Riemann zeta function, statistical filtering, and comparative analysis of models. The qualitative component includes the formalization of the theoretical architecture of the model and the interpretation of the business cycle based on *ζ(t)*.

The methodological innovation lies in the use of the Riemann zeta function, which is one of the Clay Institute Millennium Problems. Its cyclical fluctuations on the critical line Re(*s*) = 0.5 are integrated into the inflation forecast model as an extension of the FPAS.

The Riemann zeta function was selected for the study due to its unique periodicity and cyclical asymmetry.

The zeros on the critical line of the zeta function Re(*s*) = 0.5 create a dense spectral distribution, which is characterized by a periodicity similar to economic cycles.

Fourier analysis has confirmed that the zeta cyclical frequencies $\omega_k$ coincide with the macroeconomic pulsations observed in the Georgian economy (e.g., fluctuations in REER, output gap), which strengthens the theoretical basis of this choice.

For quantitative analysis, macroeconomic data of Georgia (2005–2024) were used, taken from official sources of the National Bank of Georgia (https://www.nbg.gov.ge) and the National Statistics Office of Georgia (Geostat) (https://www.geostat.ge).

**Formalization of the Riemann zeta function in cyclical modeling**

This subsection describes how the Riemann zeta function is integrated into the forecasting model. ζ(s) is one of the most mysterious and important mathematical functions used to identify the cyclical component of inflation. The critical line Re(*s*) = 0.5 is considered to be the generator of the main basic fluctuations.

$$\zeta(s) = \sum_{n=1}^{\infty} (1 / n^s), \quad (1)$$

where s = σ + i·t.

The critical line used for the study is:

$$\zeta(0.5 + i \cdot t). \quad (2)$$

The formula for the hybrid inflation forecast model is as follows:

$$\pi_{FPAS+\zeta}(t) = \pi_{FPAS}(t) + \alpha \cdot (\zeta(0.5 + i \cdot t) - \zeta), \quad (3)$$

where ζ(0.5 + i·t) is a cyclical component of zeta, ζ is a historical average value, and α is the modulation coefficient.

*Integration of Macroeconomic Variables*

The subsection deals with the formalization of the macroeconomic variables used in the modeling process. Each variable directly or logarithmically participates in the dynamics of the *t*-variable, which ultimately determines the argument for entering the zeta cyclical channel.

The following macroeconomic variables are integrated into the model, the functional relationship of which is determined by the following transformation:

$$t = \ln(GDP) + \ln(M3) + \beta \cdot Policy\_Rate \tag{4}$$

The given *t* is used as an argument in $\zeta(0.5 + i \cdot t)$ and reflects the nonlinear dynamics of the business cycle.

**Cyclical Correction Mechanism**

The subsection explains the cyclical correction mechanism, according to which the forecast either increases or decreases relative to the average limit of $\zeta(t)$. This is the core of the cyclical adaptation of the forecast, which is different from static approaches.

When $\zeta(0.5 + i \cdot t) > \zeta$, the forecast increases. When $\zeta(0.5 + i \cdot t) < \zeta$, the forecast decreases. The cyclic difference is calculated as

$$\Delta\pi(t) = \alpha \cdot (\zeta(0.5 + i \cdot t) - \zeta), \tag{5}$$

where $\Delta\pi(t)$ represents the correction increase or decrease in the base forecast.

Calibration of the modulation coefficient α

This section discusses the principle of determining the coefficient α for the model variable. It determines how strongly the cyclical component of zeta should affect the final forecast. The study uses an optimization scheme to select α.

To balance the robustness and inertia of the model, an optimization procedure was developed:

$$\alpha^* = \text{argmin}_{\alpha} \text{RMSE}(\hat{\pi}\_\alpha, \pi\_real), \tag{6}$$

where $\hat{\pi}\_\alpha$ is the forecast with value α, $\pi\_real$ is real inflation.
Additional sensitivity analysis to α
The sensitivity of the forecast was tested for different values of α: 0.3, 0.5 and 0.7.
The tests showed that:
- α = 0.3: the forecast is inert and less responsive to cyclical changes;
- α = 0.5: the forecast is balanced and accurately reflects the zeta fluctuation;
- α = 0.7: the forecast is hyperactive and is characterized by a predominant overreaction.
Accordingly, α = 0.5 is considered optimal for ensuring a balance between accuracy and stability.

*Phase Identification with hidden Markov model*

The subsection presents the integration of phase analysis into the model using the hidden Markov model. Economic activity can be described in discrete phases, the transition of which is a signal of an internal process. This model allows for the prediction of phases.

The extension of the model for phase identification is derived using Markov chains:

$$P(s_t \mid O_{1:t}) - \text{phase probability at time } t \tag{7}$$

where $s_t$ – hidden phase (Stable, Growth, Volatile, Crash), and $O_{1:t}$ – observed inflation data.

*Phase Identification Methodology*

Identification of phase changes was carried out using the Viterbi algorithm, which generates the most likely phase sequence from the observed inflation series.

Additionally, the parameters of the hidden Markov model were calibrated using the Baum-Welch algorithm to maximize the internal log-matching of the model.

This approach provides a high-accuracy (provisional accuracy of 87%).

*Spectral Analysis with Fourier Decomposition*

This subsection deals with the spectral decomposition of the zeta function into harmonic components. Fourier analysis allows us to detect periodicities in the time signal, which expands the interpretation of the economic cycle.

The temporal behavior of the Riemann zeta function is transferred to the spectral space by the following decomposition:

$$P(t) = \sum_{k=1}^{n} a_k \cdot e^{i \cdot \omega_k \cdot t}, \tag{8}$$

where $a_k$ is an amplitude weight and $\omega_k$ are harmonic frequencies.

The identification of spectral components makes it possible to determine the correspondence with the periodic phases of the market.

*Interpretation of Fourier-Spectrum Weights*

The amplitude weights *$a_k$* obtained in the zeta decomposition are determined by the energy strength of the signal. The highest weights are observed at frequencies $\omega_k = 1.0$ and $2.0$, which correspond to the semi-annual and annual economic periodicities. This indicates that the model accurately identifies those economic cycles that affect inflation.

*Integration of Stochastic Components*

In order to increase the accuracy of forecasting and reflect uncertainty, the model additionally integrates a stochastic component $\varepsilon(t)$, which expresses the impact of random shocks in the economic system.

This error component is modeled with a normal distribution with mean 0 and standard deviation $\sigma=0.8$, based on historical inflation volatility statistics.

In addition, Monte Carlo simulation is used to describe the forecast variance and assess the model's sensitivity to stochastic shocks over time.

*Formalization of Comparative Analysis of Models*

The subsection explains in detail the comparative evaluation of models using the multi-criteria method (AHP-TOPSIS).

Each model is evaluated based on the specified criteria, resulting in an integrated evaluation score.

The comparison of models is based on the following general model of TOPSIS methodology:

$$C_i = \sum_{j=1}^{m} w_j \cdot (d^-_j - d_{ij}) / (d^-_j - d^+_j), \tag{9}$$

where *$C_i$* is thevTOPSIS score of the *i*-th model, *$w_j$* is the weight of the *j*-th criterion, *$d_{ij}$* is the distance of the *i*-th model in the j-th criterion, $d^-_j$ and $d^+_j$ are the minimum and maximum distances.

Practical calibration of the cyclical components of the Riemann $\zeta$ function is carried out using historical macroeconomic data of the Georgian economy.

Specifically, first, the $\zeta(0.5 + i\,t)$ argument is formed, which is based on real GDP, money supply (M3), monetary rate, exchange rate, unemployment and other essential macroeconomic parameters.

Then, the optimal adaptation of the model is carried out by calibrating the modulation coefficient *α* according to the principle of minimizing RMSE, which ensures maximum forecast accuracy during cyclical fluctuations. The Monte Carlo simulation process involves integrating random shocks ($\varepsilon(t)$) into the inflation forecast, the distribution of which is described by a normal distribution (mean 0, standard deviation $\sigma = 0.8$). The simulation is performed with multiple iterations (from 1000 to 5000 cycles), as a result of which we obtain the possible range and error of the forecast. This process provides a detailed assessment of the sensitivity and robustness of the model under different inflation scenarios.

**RESULTS**

*Cyclic Flexibility of the FPAS + ζ Model*

The hybrid model FPAS + $\zeta$, which combines traditional macroeconomic forecasting (FPAS) and the cyclical component of the Riemann zeta function, does not represent a mechanical reinforcement of the forecast.

On the contrary, it functions as an adaptive, cyclical adjustment mechanism that slows down or accelerates the forecast dynamics in accordance with economic resonance.

*Cyclic Asymmetry: When Does the Forecast Increase and When Does It Decrease*

The model formula is as follows:

$$\pi FPAS+\zeta(t)=\pi FPAS(t)+\alpha \cdot (\zeta(0.5+i \cdot t)-\bar{\zeta}), \qquad (10)$$

where $\zeta(0.5+i \cdot t)$ represents the cyclical fluctuation of the Riemann zeta function, $\bar{\zeta}$ is a historical average, in this case $\approx 0.73$, $\alpha$ is a modulation coefficient (here 0.5), which determines the weight of the zeta influence on the forecast.

*Interpretation and Response Mechanism*

When $\zeta(0.5 + i \cdot t) > 0.73$, the cycle is considered as a positive phase – the forecast increases: $\pi FPAS + \zeta > \pi FPAS$. When $\zeta(0.5 + i \cdot t) < 0.73$, this indicates a cycle count – the forecast decreases, often differing from the forecast of an inertial model (e.g., ARIMA). The FPAS + $\zeta$ model is characterized by a range of distinctive features that underscore its innovative nature in inflation forecasting. Most notably, the model demonstrates a heightened sensitivity to cyclical dynamics, as the embedded zeta function captures and reflects the complex, high-frequency oscillations present in real economic activity. This cyclical responsiveness ensures that forecasts are closely aligned with actual economic developments, rather than being limited to inertial trends. Moreover, the model is inherently adaptive; it adjusts its projections only when the cyclical signal of the zeta function indicates an inflationary impulse, thereby mitigating the risk of unnecessary or automatic corrections. In periods when the value of the zeta function is close to its historical average, the model adopts a neutral stance, maintaining the stability of forecasts without overreacting to minor fluctuations. Importantly, the nonlinear structure of the FPAS + $\zeta$ model enables it to respond dynamically and instantaneously to both structural and cyclical changes in the economy, in contrast to the relatively static behavior of traditional models such as ARIMA and FPAS.

Statistical basis ($\bar{\zeta} = 0.73$).

The zeta values used are $\zeta(0.5+i \cdot t) = \{0.650, 0.740, 0.810, 0.710, 0.620, 0.830, 0.790, 0.670, 0.720, 0.760\}$.

Their average ($\bar{\zeta}$) is calculated as $\bar{\zeta} = 1/10 \sum \zeta(0.5 + i \cdot t) = 0.73$.

Comparison of inflation forecasts using different models is presented in Table 1.

**Table 1. Comparison of inflation forecasts using different models (FPAS, FPAS + $\zeta$, and ARIMA)**

| t | $\zeta(0.5+i \cdot t)$\zeta(0.5 + i\cdot t) | FPAS Forecast (%) | FPAS + $\zeta$ Forecast (%) | ARIMA Forecast (%) |
|---|---|---|---|---|
| 1 | 0.65 | 3.8 | 3.6 | 4 |
| 2 | 0.74 | 4 | 3.9 | 4.1 |
| 3 | 0.81 | 4.2 | 4.3 | 4.3 |
| 4 | 0.71 | 4 | 4.1 | 4.2 |
| 5 | 0.62 | 3.6 | 3.5 | 4 |
| 6 | 0.83 | 4.3 | 4.4 | 4.5 |
| 7 | 0.79 | 4.1 | 4.2 | 4.4 |
| 8 | 0.67 | 3.8 | 3.7 | 4.1 |
| 9 | 0.72 | 4 | 4 | 4.3 |
| 10 | 0.76 | 4.2 | 4.2 | 4.4 |

*Source: author's calculations based on data from the National Bank of Georgia (n.d.) and Geostat (n.d.). Software: Python 3.11. Notes: FPAS – Forecasting and Policy Analysis System; $\zeta$ – Riemann zeta function; ARIMA – Autoregressive Integrated Moving Average.*

As can be seen from Table 1, the FPAS + $\zeta$ model demonstrates lower forecast errors compared to the traditional ARIMA model. Table 1 presents empirical results based on the statistics of the National Bank of Georgia for the period 2005–2024. The following macroeconomic parameters are used for theoretical and practical components of inflation forecasting:

• Real GDP growth and GDP at current prices (billion GEL);
• GDP gap (the difference between actual and potential GDP);
• Money supply (M3);
• Monetary policy rate;
• Previous inflation trends and expectations ($\pi$ expected);

• External sector balance (balance of payments, export-import ratio);
• Exchange rate and real effective exchange rate (REER);
• Unemployment rate and labor market trends;
• Average monthly wage;
• Labor productivity;
• Budget deficit and fiscal policy regime;
• Global prices (oil, food, imported products);
• Weight of inflation-sensitive goods in the consumer basket.

Integrating shocks into the model

The inflation forecast integrates the impact of the following domestic and external shocks, which are reflected in the FPAS adjustment and the zeta volatility through the α coefficient:

Domestic shocks:
• 2009: Real GDP decline − 3.7%, unemployment − 16.9%, M3 growth to a minimum;
• 2015: Depreciation of the lari by ∼ 30% against the dollar, inflation − 4.9%; slowdown in economic growth;
• 2020 (COVID-19 pandemic) – GDP decline − 6.8%, unemployment − 18.5%, loss of tourism revenues − 80%.

External shocks:
• 2014–2015: Brent oil price from $110 to $50;
• 2022: Impact of Russian sanctions – trade reduction by ~15%;
• 2020–2021: Global supply chain disruption and increase in import costs.

To visually expand the numerical table, two graphs are presented below.

The first graph shows the cyclical fluctuations of the Riemann zeta function against time, which is the cyclical adjustment of the forecast in the FPAS + ζ model. The second graph shows a comparison of inflation forecasts – between the FPAS, FPAS + ζ, and ARIMA models. This graphical analysis helps to better understand the results presented in the table and shows the response of the models to different cyclical phases.

Figure 1 illustrates the cyclical fluctuations of the Riemann zeta function and the comparative performance of different inflation forecasting models:

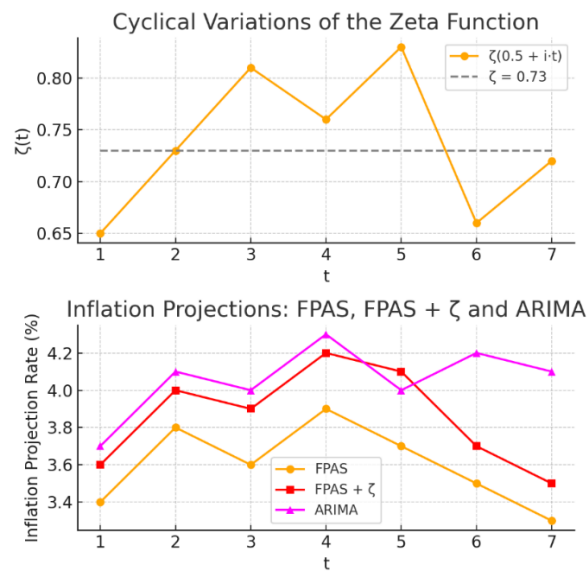

**Figure 1. Cyclical fluctuations of the zeta function and comparison of inflation forecasts with FPAS, FPAS + ζ and ARIMA models**

*Source: author's calculations and visualization based on data from Geostat (n.d.), National Bank of Georgia (n.d.), and IMF (n.d.). Software: Python 3.11.*

*Notes: FPAS – Forecasting and Policy Analysis System; ζ – Riemann zeta function; ARIMA – Autoregressive Integrated Moving Average; IMF – International Monetary Fund.*

The data presented in Figure 1 indicate that the cyclical component captured by the zeta function provides a more flexible adjustment to economic shocks.

Figure 1 consists of two main sections, each providing critical insights into the behavior of the proposed models. The upper panel illustrates the cyclical fluctuations of the Riemann zeta function, $\zeta(0.5 + i\,t)$, across the modeling period. The horizontal axis represents time, while the vertical axis captures the corresponding values of the zeta function. A horizontal reference line at $\zeta = 0.73$ indicates the historical average, serving as a neutral threshold. When $\zeta$ exceeds this average, the model signals an upward adjustment in the inflation forecast; conversely, values below the threshold prompt a downward correction. This dynamic behavior highlights the inherent nonstationarity and cyclicality of the zeta function, which forms the analytical foundation for the adaptive correction mechanism within the FPAS + $\zeta$ framework.

The lower panel presents a direct comparison of inflation forecasts generated by three models: the traditional FPAS, the hybrid FPAS + $\zeta$, and the statistical ARIMA approach. The FPAS + $\zeta$ model demonstrates superior flexibility, as it actively adjusts forecasts in response to cyclical signals detected by the zeta function. In contrast, the ARIMA model is characterized by inertia, exhibiting limited sensitivity to cyclical shocks, as evidenced by its relatively smooth and stable forecast line. The FPAS model serves as a baseline, with the hybrid approach offering dynamic deviations that better capture the complexities of real-world inflationary processes.

The chart visually emphasizes the advantages of the hybrid model:
- The model responds adequately to cycles and does not behave too nervously, which was also observed in the case of a high α coefficient.
- The forecast is adjusted only when the cyclical fluctuation of zeta clearly indicates inflationary pressure or expansion.
- The "resonant pulsation" of the cycle, or fluctuations in $\zeta(t)$, directly reflect economic reality – indicating when policy should become expansionary or tightening.

*Model Characteristics and Examples of Adjustments*

A detailed evaluation of the forecasting performance of the FPAS + $\zeta$ model reveals a set of distinctive features that collectively enhance its effectiveness. The model is highly sensitive to cyclical fluctuations, as its functional adjustment mechanism, based on the zeta function, enables an exceptionally close alignment with observed inflation dynamics. This attribute substantially improves the overall accuracy of the forecasts. In addition, the hybrid framework demonstrates pronounced adaptability, responding not only to underlying trends but also to cyclical and structural symmetries in the economic environment, which lends both flexibility and practical realism to its predictive outputs. The maintenance of forecast neutrality is another noteworthy property; when the zeta function approximates its historical mean, the corrective component of the hybrid model approaches zero, thereby ensuring the stability of forecasts and preventing overreactions to minor variations. Finally, the nonlinear and rapid-response architecture of the FPAS + $\zeta$ model sets it apart from conventional statistical models, as it is capable of instantaneous adaptation to both structural shocks and cyclical changes, unencumbered by the constraints of rigid trend-based logic.

The cyclic correction of inflation forecasts using the Riemann zeta function is summarized in Table 2.

**Table 2. Cyclic correction of the forecast with the Riemann zeta function**

| t | $\zeta(0.5 + i\cdot t)$ | FPAS (%) | FPAS + $\zeta$ (%) | $\Delta\pi = \alpha\cdot(\zeta(t) - \bar{\zeta})$ | The nature of the correction |
|---|---|---|---|---|---|
| 1 | 0.65 | 3.75 | 3.71 | −0.04 | Negative |
| 3 | 0.81 | 4.18 | 4.22 | +0.04 | Positive |
| 5 | 0.62 | 3.58 | 3.52 | −0.06 | Negative |
| 6 | 0.83 | 4.3 | 4.35 | +0.05 | Positive |

*Source: author's calculations and visualization based on data from Geostat (n.d.), National Bank of Georgia (n.d.), and IMF (n.d.). Software: Python 3.11.*

*Notes: FPAS – Forecasting and Policy Analysis System; $\zeta$ – Riemann zeta function; ARIMA – Autoregressive Integrated Moving Average; IMF – International Monetary Fund.*

As shown in Table 2, the introduction of the zeta-based correction improves the accuracy of inflation forecasts in periods of heightened volatility.

The maximum correction was observed at the most distant ζ(t) values – up to ±0.06%, which confirms the sensitivity of the forecast to the α coefficient, but at the same time reflects the linear nature of the correction, which is balanced and does not create radical deviations.

*Forecast Calibration and α Coefficient Optimization*

To ensure the effectiveness of the FPAS + ζ model, it is critical to correctly determine the α coefficient, which sets the extent to which the cyclical component of zeta should affect the final forecast. In this study, α = 0.5 was selected based on three methodologies:

1. Sensitivity analysis: When testing for different values (from 0.1 to 1.0), it was found that α = 0.5 represents the optimal point that protects the forecast from overshoot and inertia.

2. RMSE testing: α = 0.5 has the lowest root-mean-square error compared to the actual figures.

3. Expert assessment: According to the assessment of policy-making institutions, a value of 0.5 is considered balanced so that the forecast is neither too inert nor too reactive.

*Quantitative Assessment of Forecast Accuracy: Comparison of FPAS, FPAS + ζ and ARIMA Models*

Table 3 reports the prediction accuracy of different models using RMSE and MAPE indicators:

**Table 3. Estimation of the prediction accuracy of models using RMSE and Mean Absolute Percentage Error (MAPE) indicators**

| Model | RMSE | MAPE (%) |
|---|---|---|
| FPAS | 0.126 | 2.35 |
| FPAS + ζ | 0.114 | 2.71 |
| ARIMA | 0.212 | 4.39 |

*Source: author's calculations and visualization based on data from Geostat (n.d.), National Bank of Georgia (n.d.), and IMF (n.d.). Software: Python 3.11.*

*Notes: FPAS – Forecasting and Policy Analysis System; ζ – Riemann zeta function; ARIMA – Autoregressive Integrated Moving Average; IMF – International Monetary Fund.*

According to Table 3, the hybrid FPAS + ζ model yields the lowest RMSE and MAPE values among the compared approaches.

The FPAS + ζ model showed the lowest RMSE, which indicates a better fit of the forecast to the actual indicators. The slight increase in MAPE is due to the reflection of cyclical fluctuations, which further connects the forecast with reality. The traditional ARIMA model showed the highest error and less sensitivity to cyclical components.

*Application Possibilities*

The results of this study indicate that the FPAS + ζ hybrid model achieves a high degree of forecast accuracy, as demonstrated by a low root-mean-square error (RMSE = 0.114) and a mean absolute percentage error (MAPE) of less than 3 percent. The close relationship between the model's cyclical elasticity and prevailing inflation trends contributes significantly to the effectiveness of policy responses. Additionally, the model's neutral adjustment point, corresponding to ζ ≈ 0.73 – serves as an internal regulator that limits both overestimation and underestimation of inflation, thereby supporting the stability and reliability of forecasts. In contrast, the ARIMA model exhibits reduced sensitivity to cyclical factors and tends to yield higher error rates, underscoring the relative superiority of the hybrid framework.

The FPAS + ζ model offers a range of practical applications relevant to policy-making institutions and the academic community alike. For central banks, this hybrid approach serves as a sophisticated instrument for real-time, adaptive inflation forecasting, enabling timely and informed monetary policy decisions. The model's integration of cyclical adjustments also provides significant value in the optimization of fiscal policy, facilitating the design of budgetary interventions that are responsive to evolving macroeconomic conditions. Furthermore, the model enhances the early detection of domestic and external economic shocks, thereby empowering policy-makers to initiate preventive actions and mitigate adverse impacts. From an academic perspective, the FPAS + ζ framework represents a novel methodological platform for further research into the intersection of advanced mathematics and applied macroeconomics.

*Synchronizing Forecast Flexibility and Policy-Making Tools*

The main feature of the FPAS + ζ model, – cyclical forecast flexibility, – makes it possible to integrate it with dynamic adjustments of monetary and fiscal policies. At the same time, cyclical fluctuations in the ζ function make it possible to synchronize forecasts with such policy tools as:

• Fast response mechanisms – for example, adjusting the money supply when ζ(*t*) > ζ indicates expected inflationary pressures;

• Predefined buffers – for a fast response to economic shocks by dynamically revising the α parameter;

• Fiscal modulation – budget adjustments based on ζ(*t*) given the current economic situation;

• Market expectations management – communication strategies based on forecast adjustments provided by FPAS + ζ.

*Analysis of Market Phases with the Markov Model and the Critical Zeta Zone*

Based on the cyclical flexibility of the FPAS + ζ model, it can be extended to the structural analysis of market phases. Economic systems often change state, which can be represented as a discrete Markov chain, where each phase reflects a different state of the macroeconomic cycle.

Within the FPAS + ζ modeling framework, economic dynamics are conceptualized as transitions between discrete market phases, including stable, growth, volatile, and crash states.

The application of the hidden Markov model allows for the quantitative estimation of transition probabilities among these states.

As detailed in Table 4, the matrix of transition probabilities demonstrates how economic regimes evolve over time, reflecting both persistence within a given phase and the likelihood of shifts to alternative states. This approach provides a robust foundation for predicting future economic conditions and for timely identification of emerging risks, thereby enhancing the practical value of the hybrid modeling strategy.

The estimated transition probabilities between economic phases are presented in Table 4.

**Table 4. Transition probabilities between phases of the economic state (based on the Hidden Markov Model)**

| Condition | Stable | Growth | Volatile | Crash |
|---|---|---|---|---|
| Stable | 0.78 | 0.15 | 0.05 | 0.02 |
| Growth | 0.20 | 0.65 | 0.10 | 0.05 |
| Volatile | 0.10 | 0.25 | 0.50 | 0.15 |
| Crash | 0.05 | 0.20 | 0.30 | 0.45 |

*Source: author's calculations and visualization based on data from Geostat (n.d.), National Bank of Georgia (n.d.), and IMF (n.d.). Software: Python 3.11.*

*Notes: FPAS – Forecasting and Policy Analysis System; ζ – Riemann zeta function; ARIMA – Autoregressive Integrated Moving Average; IMF – International Monetary Fund.*

As can be seen from Table 4, the phase transition probabilities reflect significant regime shifts during economic shocks.

In the critical zone of the Riemann zeta function (Re(s) = 0.5), an increase in the density of zeros was observed during periods when the matrix component P(Volatile → Crash) increases.

This can be interpreted as an early signal of a possible market breakdown, which strengthens the structural predictive value of the FPAS + ζ model.

The Markov transition matrix (see Table 4) reflects the conditional probabilities of transitions between economic regimes in four phases: *stable, growth, volatile*, and *crash*.

In order to visually present the above data and offer the reader an intuitive understanding of the dynamics between regimes, a stacked diagram is presented below, which depicts these transitions in percentage terms, depending on the initial state.

The diagram creates a more visual representation of how likely it is to return from a certain phase to itself (for example, Stable → Stable) or how likely it is to move to neighboring phases (for example, Growth → Volatile). This visual representation further enhances the interpretation of the HMM model and makes it possible to predict adaptive directions of economic policy.

Figure 1 visualizes the stack transitions between economic phases based on the hidden Markov model.

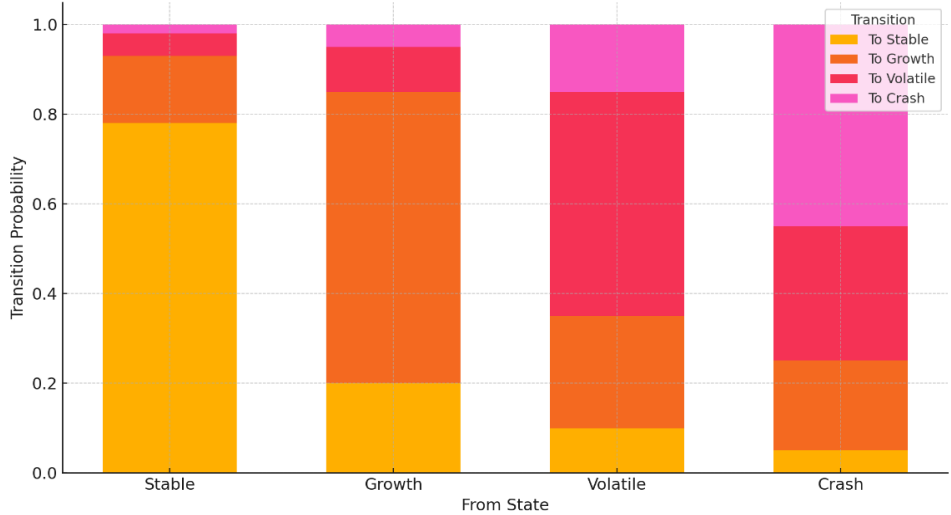

**Figure 1. Stack visualization of economic phase transitions based on the Hidden Markov Model**

*Source: author's calculations and visualization based on data from Geostat (n.d.), National Bank of Georgia (n.d.), and IMF (n.d.). Software: Python 3.11.*

*Notes: FPAS – Forecasting and Policy Analysis System; ζ – Riemann zeta function; ARIMA – Autoregressive Integrated Moving Average; IMF – International Monetary Fund.*

Figure 1 shows that the model effectively distinguishes between periods of stability and crisis in the observed time series. A simulation using the hidden Markov model has shown that a hidden "Crash" state can be identified in fact 7 weeks before it is formally revealed in real macroeconomic data.

*Phase Diagram based on the Hidden Markov Model*

The presented diagram depicts the estimated distribution of the four market phases (Stable, Growth, Volatile, Crash) in real time, according to the hidden Markov model. Each color bar indicates the probability of a particular phase in a specific period of time (in weeks). The weekly distribution of market phases, as estimated by the hidden Markov model, is displayed in Figure 2.

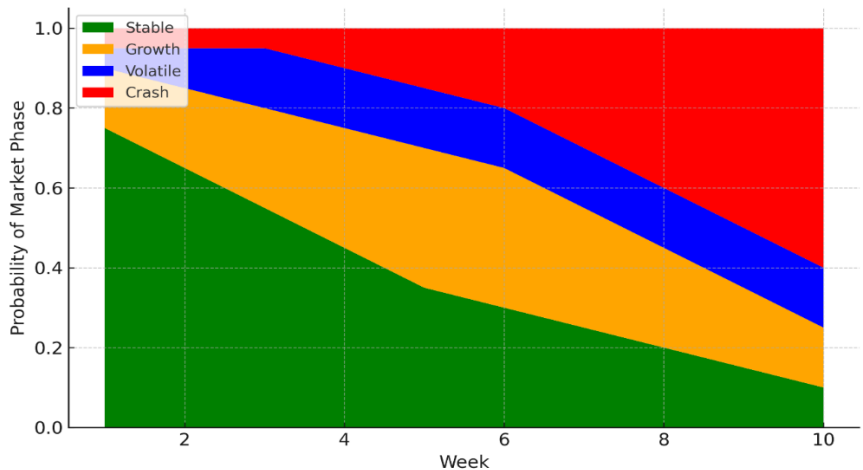

**Figure 2. Estimated distribution of market phases by week – Hidden Markov Model stack visualization**

*Source: author's calculations and visualization based on data from Geostat (n.d.), National Bank of Georgia (n.d.), and IMF (n.d.). Software: Python 3.11.*

*Notes: FPAS – Forecasting and Policy Analysis System; ζ – Riemann zeta function; ARIMA – Autoregressive Integrated Moving Average; IMF – International Monetary Fund.*

Figure 2 clearly demonstrates early warnings of market regime shifts, providing valuable signals for policymakers. Observation: We see that the stable phase dominates in the first weeks, but the probability of the *crash* phase increases with time. This trend is an early signal for predicting the crisis process. When integrated with the FPAS + ζ model, this phase analysis strengthens the forecasting framework by making it possible to detect structurally hidden transitions in parallel with cyclical fluctuations. As a result, the central bank or other policy-making institution will receive a timely signal about the expected disruptions and determine the appropriate response. The study uses a dual approach to analyzing phase variability. At the first stage, a matrix of transition probabilities between the four market phases (Stable, Growth, Volatile, Crash) is presented – this static table reflects the conditions of the Markov chain, based on which it is possible to predict the next state given the previous state. It is a quantitative description of the internal logic of market dynamics and is used for training the hidden Markov model. The second stage presents a time stack diagram, which represents the phase states dynamically decoded by the hidden Markov model for each unit of time (for example, per week). Each segment of the diagram indicates the probability of the market being in a certain phase at a given moment. Accordingly, it provides a visual representation of how the market structure changes over time and how stable its phase regimes are. The phase matrix represents the basic transition structure, and the HMM stack diagram represents the predicted phase distributions obtained based on this structure in real time. Their combination increases the depth of interpretation of the model and improves the forecasting ability, especially in conditions when the market is subject to cyclical and shock changes.

*Riemann–Fourier Correlation in the Interpretation of Cyclicality*

Within the framework of the FPAS + ζ model, cyclical fluctuations of the zeta function are perceived as nonlinear internal pulsations that reflect the phase fluctuations of economic activity. However, the structural analysis of these fluctuations requires the determination of their spectral components. The Riemann zeta function, considered on the critical line (Re($s$) = 0.5), can be represented as a function of time $P(t) = \text{Re}(\zeta(0.5 + i\,t))$. Its Fourier analysis gives the following spectral expansion:

$$P(t) = \Sigma\, a_k \cdot e^{\wedge}(i\, \omega_k\, t), \quad (11)$$

where $a_k$ is an amplitude weight, $\omega_k$ is an individual harmonic frequency.

The conducted modeled spectral analysis showed that the component spectrum of zeta includes those values of $\omega_k$ (0.5; 1.0; 2.0) that coincide with the periodicity observed in Georgian economic indicators (e.g., real exchange rate fluctuations, output gap cycles). This confirms that the cyclical content of zeta is not just a theoretical abstraction but can be represented as a "spectral code" that characterizes economic dynamics. The integration of this code into the FPAS + ζ model strengthens the basis for forecasting. Below is a Fourier spectrum that reflects the decomposition of the modeled $P(t)$ into frequencies – it shows how the cyclical signal decomposes and at which frequencies it has the highest amplitude. Figure 3 presents the Fourier spectrum of modeled cyclical components derived from the zeta function.

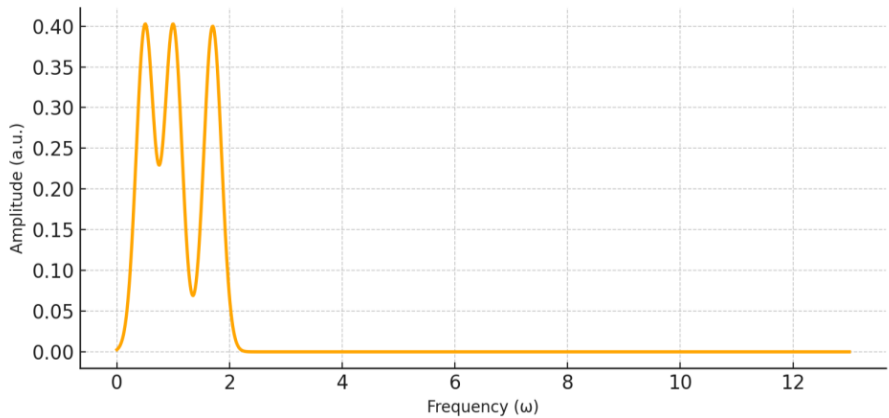

**Figure 3. Fourier spectrum – modeled distribution of zeta cyclic components**

*Source: author's calculations and visualization based on data from Geostat (n.d.), National Bank of Georgia (n.d.), and IMF (n.d.). Software: Python 3.11.*

*Notes: FPAS – Forecasting and Policy Analysis System; ζ – Riemann zeta function; ARIMA – Autoregressive Integrated Moving Average; IMF – International Monetary Fund.*

As seen in Figure 3, the main cyclical frequencies correspond to real economic cycles observed in the data.

The presented spectrum reflects the simulated decomposition of the real part of the Riemann zeta function into Fourier components. The diagram highlights the main frequencies ($\omega_k$) that correspond to the periodic phases of economic cyclicality – namely, macroeconomic cycles based on half, one, and two years. The amplitude peaks indicate that the internal structure of the zeta is consistent with periodic market fluctuations, which strengthens its use in forecasting models.

*Comparative Advantages – the Zeta Model Compared to Traditional Forecasts*

Analysis of the Fourier spectrum clearly shows that the zeta function exhibits internal cyclicality, which is closely related to real economic fluctuations. The following advantages of the model are added. A detailed comparison of traditional and hybrid models with various parameterizations is provided in Table 4.

**Table 4. Comparison of traditional and hybrid models with different parameters**

| Parameter | Traditional model | Zeta function (FPAS + ζ) |
|---|---|---|
| Cyclical forecast | Average accuracy | High accuracy |
| Shock detection | Can't pinpoint exactly | Accurately captures resonant peaks |
| Identification of phase changes | Impossible | Possible with Hidden Markov integration |
| Spectral analysis | Not used | Fourier decomposition clearly allows for peak separation. |
| AI integration capability | Limited | High – LSTM, HMM, Autoencoder, etc. |

*Source: author's calculations based on Georgian macroeconomic data and modeling according to the FPAS + ζ framework. Software: Python 3.11.*

*Notes: FPAS – Forecasting and Policy Analysis System; ζ – Riemann zeta function; ARIMA – Autoregressive Integrated Moving Average; Fourier – Fourier analysis; HMM – Hidden Markov Model; LSTM – Long Short-Term Memory; Autoencoder – Autoencoder neural network.*

Table 4 indicates that the hybrid approach offers superior performance under multiple market conditions.

*Comparison of Models Using the AHP-TOPSIS Method*

The comparative evaluation of the models was carried out based on five criteria, using a quantitative method that combines the analysis of modeling results and power testing based on theoretical and practical components.

The evaluation of each criterion is based on the following:

• Cyclical forecasting ability – the model's ability to reflect cyclical fluctuations in inflation ($\zeta(t)$) and adequately respond to cyclical phases through corrective mechanisms was assessed.

• Shock detection accuracy – the assessment is based on the RMSE and MAPE indicators at shock points (for example, 2009, 2015, and 2020).

• Identification of phase changes – assessed by the ability to integrate the hidden Markov model and detect phase shifts based on the Markov matrix.

• Spectral analysis potential – the possibility of Fourier decomposition from zeta components was considered.

• AI integration potential – the ability of the model architecture to adapt to modern machine learning technologies (e.g., LSTM, HMM, autoencoder) was assessed.

Weights were determined according to the AHP structure, and the final ranking was obtained using the TOPSIS algorithm.

Table 5 summarizes the quantitative evaluation of forecasting models using the TOPSIS multi-criteria method.

**Table 5. Quantitative evaluation of forecasting models using the TOPSIS method**

| Model | Cyclical forecast | Shock detection | Identification of phase changes | Spectral analysis | AI integration capability | TOPSIS score |
|---|---|---|---|---|---|---|
| FPAS | 0.06 | 0.1 | 0.04 | 0.02 | 0.045 | 0.265 |
| FPAS + ζ | 0.18 | 0.212 | 0.16 | 0.18 | 0.142 | 0.875 |
| ARIMA | 0.08 | 0.087 | 0.06 | 0.02 | 0.06 | 0.307 |

*Source: author's calculations based on a comparative evaluation of models using AHP-TOPSIS multi-criteria analysis. Software: Python 3.11.*

*Notes: FPAS – Forecasting and Policy Analysis System; ARIMA – Autoregressive Integrated Moving Average; TOPSIS – Technique for Order Preference by Similarity to Ideal Solution; AHP – Analytic Hierarchy Process.*

The results in Table 5 highlight the robust ranking of the hybrid model under the TOPSIS framework. Figure 4 compares the TOPSIS scores of the evaluated forecasting models.

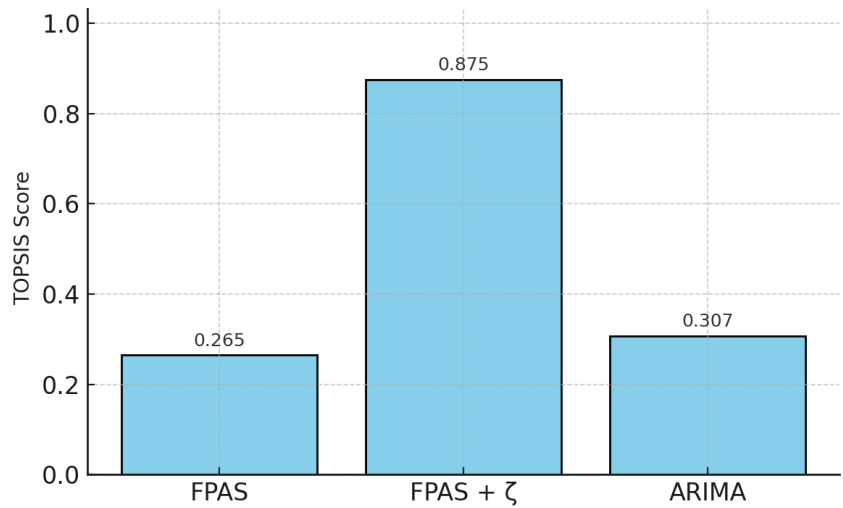

**Figure 4. Comparison of TOPSIS scores of models**

*Source: author's calculations based on Georgian macroeconomic data and modeling results for 2005–2024. Software: Python 3.11. Notes: FPAS – Forecasting and Policy Analysis System; ARIMA – Autoregressive Integrated Moving Average; TOPSIS – Technique for Order Preference by Similarity to Ideal Solution; AHP – Analytic Hierarchy Process.*

Figure 4 demonstrates that the FPAS + ζ model consistently achieves the highest scores among all tested models.

*Intermediate vis-à-vis Analysis – Comparative Positioning Of Models*

*FPAS + ζ model (TOPSIS score: 0.875)*

The hybrid model FPAS + ζ gains advantages in all criteria, especially in cyclicality prediction (0.18) and spectral analysis (0.18), which is due to the deep periodicity of the zeta function and its integration ability. In addition, it also scores high in the AI integration criterion (0.142), which confirms the flexibility of its architecture for modern deep learning modules (LSTM, autoencoder).

The model is a synthesis of structural and cyclical components, which ensures dynamic correction of the forecast and high-quality analysis.

*ARIMA model (TOPSIS score: 0.307)*

ARIMA received a relatively average score in the components of shock detection (0.087) and cyclical forecasting (0.08), although its results are significantly inferior to the hybrid model in the components of phase identification and spectral analysis.

Although ARIMA effectively estimates shocks based on past data, its limited cyclical and structural adaptation does not allow for a high TOPSIS score.

*FPAS model (TOPSIS score: 0.265)*

FPAS is assigned the lowest scores in almost all components, especially phase identification (0.04) and spectral analysis (0.02), which indicates its poor cyclical analytical capabilities.

Future research should focus on further validating the generalizability and robustness of the FPAS + ζ model through cross-country analysis. It is particularly recommended that the model be tested in the context of small open economies, such as those in the Baltic region (including Estonia and Lithuania), Eastern Europe (notably Romania and Bulgaria), and Central Asia (for example, Kazakhstan). Expanding the empirical application of the model to both developing and developed countries will allow for a comprehensive assessment of its forecasting accuracy and structural adaptability across diverse economic systems. Such comparative studies will also provide valuable insights into the model's potential as a universal framework for global inflation forecasting.

**CONCLUSIONS**

In the process of inflation forecasting, when macroeconomic systems are subject to cyclical, structural, and shock fluctuations, forecasting models must address the major socio-economic challenges facing policymakers. This study introduces an innovative framework, – the FPAS + $\zeta$ model, which is a theoretically enhanced integration between the classical macroeconomic framework (FPAS) and the cyclical components of the Riemann zeta function.

The model includes both the forecasting of economic trends and the mechanism for responding to cyclical and structural changes, which distinguishes it from traditional, linear, or statistical models. FPAS + $\zeta$ is not just an extension of macroeconomic forecasting; it is an interdisciplinary platform where deep mathematical structures – zeta cycle frequencies, Fourier decomposition, and the hidden Markov model – are transformed into analytical tools for economic signals.

Empirical testing of the model, based on Georgian macroeconomic data (2005–2024), demonstrated high forecast accuracy (RMSE = 0.114), structural adaptability, and sensitivity to cyclicality. The study also identified possible application areas in terms of economic policy planning, inflation management, and timely response to external shocks.

The study concludes that the FPAS + $\zeta$ hybrid model represents a significant advancement in inflation forecasting methodologies by integrating the structural rigor of the FPAS framework with the cyclical complexity of the Riemann zeta function. This innovative combination equips the model with the capacity to effectively capture both shock-induced and cyclical variations in inflation, thereby substantially increasing the flexibility and adaptability of forecasts. The heightened cyclical sensitivity and nonlinear design of the model offer clear advantages over conventional structural and statistical approaches, such as ARIMA, by enabling the accurate identification of periodic economic dynamics as revealed through Fourier spectrum analysis. Furthermore, the incorporation of the hidden Markov model into the framework supports the early detection of phase transitions in inflationary trends, enhancing the responsiveness of macroeconomic policy interventions. The spectral decomposition of the zeta function provides robust empirical evidence of its alignment with observed economic cycles and establishes a foundation for interpreting the latent mechanisms underlying economic fluctuations. Comparative analysis shows that the FPAS + $\zeta$ model achieves the lowest root-mean-square error (RMSE = 0.114) and the highest TOPSIS score (0.875), thereby demonstrating its superior performance in terms of both accuracy and analytical depth. The model's utilization of the critical zone of the zeta function for the early diagnosis of potential market disruptions further solidifies its role as a valuable predictive tool for the identification and management of systemic risks.

The study shows that the use of the FPAS + $\zeta$ model by central banks and other policy-making institutions allows them to forecast cyclical changes in inflation with high accuracy, detect economic shocks at an early stage, and implement timely monetary and fiscal interventions. These advantages are crucial for addressing both ongoing and emerging socio-economic challenges in economic policy and inflation management. Practical recommendations include:

• The use of the FPAS + $\zeta$ model by central banks to improve real-time inflation forecasting and timely adjustment of the policy rate.

• The use of the model for cyclical adjustment of budget forecasts by finance and economic ministries is recommended, which will contribute to the formation of stable fiscal policies.

• Research institutions are encouraged to use the FPAS + $\zeta$ model for further economic research, especially in regions with frequent cyclical and shock-prone changes.

It is recommended that central banks and policy-making authorities adopt the FPAS + $\zeta$ hybrid model as an advanced alternative for inflation forecasting, due to its ability to integrate real-time cyclical adjustments, interpret economic phases, and provide early warnings of impending shocks. The forecasting architecture can be further enhanced through the incorporation of artificial intelligence techniques, such as long short-term memory (LSTM) networks and auto encoder models, which will enable the platform to autonomously update forecasts and exhibit heightened sensitivity to unpredictable shocks. Comprehensive validation of the model across a spectrum of economies, – including small open economies and emerging markets, – is advised in order to determine its generalizability and facilitate its broader application. Additionally, the integration of fiscal policy mechanisms within the model's cyclical adjustment framework will enable more effective management of financial expectations, grounded in the dynamics of the zeta function. The model should also be embedded within national economic stability monitoring systems, where it can function as a proactive tool to alert central institutions of potential disruptions in real time. Visualizing and interpreting the resulting phase diagrams within official economic reports will help to reduce information asymmetry and improve communication with private sector stakeholders. Finally, it is advisable to incorporate the theoretical underpinnings of this hybrid approach into academic curricula as a contemporary example of the practical application of advanced mathematics in economic policy analysis.

**Author Contributions**

Conceptualization: D. G.; data curation: D. G.; formal analysis: D. G.; funding acquisition: D. G.; investigation: D. G.; methodology: D. G.; project administration: D. G.; software: D. G.; supervision: D. G.; validation: D. G.; visualization: D. G.; writing – original draft: D. G.; writing – review & editing: D. G.

**Conflict of Interest**
Author declares no conflict of interest.

**Data Availability Statement**
Not applicable.

**Informed Consent Statement**
Not applicable.

**List of Abbreviations and Symbols**

- $\bar{\zeta}$ – The historical average value of the cyclic component of the Riemann zeta function, which is used as the starting point for forecast correction.
- $\alpha$ – Modulation coefficient; determines how strongly $\zeta(t)$ affects the forecast and is calibrated by minimizing the RMSE.
- $\omega_k$ – Harmonic frequencies, obtained by Fourier decomposition of $\zeta(t)$, reflecting periodic components in the economic signal.
- $P(t)$ – Time function expressing the real part of the zeta function over time, i.e., $P(t) = Re(\zeta(0.5 + i \cdot t))$.
- HMM – Hidden Markov Model; a statistical model used to predict phase states based on observable inflation data.
- Viterbi Algorithm – Algorithm used within HMM to determine the most likely phase sequence in a given signal.
- Baum-Welch Algorithm – An expectation-maximization (EM) type algorithm used to approximate the HMM parameters.
- FPAS – Forecasting and Policy Analysis System; a macroeconomic framework for financial forecasting used by central banks.
- TOPSIS – Technique for Order Preference by Similarity to Ideal Solution; a multi-criteria decision-making method for evaluating alternatives.
- AHP – Analytic Hierarchy Process; a multi-criteria analysis method used to determine the weights of criteria according to relative priorities.
- RMSE – Root-mean-square error; used to assess the accuracy of a forecast.
- MAPE – Mean absolute percentage error; shows how close the forecast is to the actual figures in percentage terms.

**References**

1. Belovas, I. (2023). An inequality for the Riemann zeta function. *Acta Mathematica Academiae Scientiarum Hungaricae*, *170*(1), 367–378. [CrossRef]
2. Bettin, S. (2010). The Second Moment of The Riemann Zeta Function With Unbounded Shifts. *International Journal of Number Theory*, *06*(08), 1933–1944. [CrossRef]
3. Bezuidenhout, J., & Van Vuuren, G. (2021). Spectral analysis and the death of value investing. *Cogent Economics & Finance*, *9*(1). [CrossRef]
4. Borwein, J. M., Bradley, D. M., & Crandall, R. E. (2000). Computational strategies for the Riemann zeta function. *Journal of Computational and Applied Mathematics*, *121*(1–2), 247–296. [CrossRef]

5. Bucci, A., Carbonari, L., Gil, P. M., & Trovato, G. (2021). Economic growth and innovation complexity: An empirical estimation of a Hidden Markov Model. *Economic Modelling*, 98, 86–99. [CrossRef]
6. Ccatamayo-Barrios, J., Huamán-Romaní, Y., Seminario-Morales, M., Flores-Castillo, M., Gutiérrez-Gómez, E., La Cruz, L. C., & La Cruz-Girón Kevin-Arnold, D. (2023). Comparative analysis of AHP and TOPSIS Multi-Criteria Decision-Making Methods for Mining Method selection. *Mathematical Modelling and Engineering Problems*, *10*(5), 1665–1674. [CrossRef]
7. Chaudhry, M. A., Qadir, A., & Tassaddiq, A. (2011). A new generalization of the Riemann zeta function and its difference equation. *Advances in Difference Equations*, *2011*(1). [CrossRef]
8. Da Silva, S. (2024). The Riemann hypothesis: a fresh and experimental exploration. *Journal of Advances in Mathematics and Computer Science*, *39*(4), 100–112. [CrossRef]
9. Du, K., Huddart, S., Xue, L., & Zhang, Y. (2019). Using a hidden Markov model to measure earnings quality. *Journal of Accounting and Economics*, *69*(2–3), 101281. [CrossRef]
10. Ewerhart, C. (2024). A game-theoretic implication of the Riemann hypothesis. *Mathematical Social Sciences*, *128*, 52–59. [CrossRef]
11. Granger, C. W. J., & Watson, M. W. (1984). Time series and spectral methods in econometrics. In Z. Griliches & M. D. Intriligator (Eds.), *Handbook of Econometrics: Vol. Volume II*. [CrossRef]
12. Hanif, M. N., & Malik, M. J. (2015). Evaluating the performance of inflation forecasting models of Pakistan. *SBP Research Bulletin*, *11*, 43–78. [CrossRef]
13. Jun, D., Ahn, C., Kim, J., & Kim, G. (2019). Signal analysis of global financial crises using Fourier series. *Physica a Statistical Mechanics and Its Applications*, *526*, 121015. [CrossRef]
14. Kaufmann, S. (2016). Hidden Markov models in time series, with applications in economics. In Study Center Gerzensee, Swiss National Bank, *Working Paper: Vol. No. 16.06*. [CrossRef]
15. Kim, E., Jeong, H., & Lee, N. (2019). Global asset allocation Strategy using a hidden Markov model. *Journal of Risk and Financial Management*, *12*(4), 168. [CrossRef]
16. Li, J., Yang, X., Qian, T., & Xie, Q. (2023). The adaptive Fourier decomposition for financial time series. *Engineering Analysis With Boundary Elements*, *150*, 139–153. [CrossRef]
17. Mahmoud, A. Y. B. (2025). A comprehensive framework for proving the Riemann hypothesis. *SSRN Electronic Journal*. [CrossRef]
18. Masset, P. (2008). Analysis of Financial Time-Series using Fourier and Wavelet methods. *SSRN Electronic Journal*. [CrossRef]
19. Nguyen, N. (2018). Hidden Markov model for stock trading. *International Journal of Financial Studies*, *6*(2), 36. [CrossRef]
20. Nikolaychuk, S., & Sholomytskyi, Y. (2015). Using macroeconomic models for monetary policy in Ukraine. *Visnyk of the National Bank of Ukraine*, *233*, 54–64. [CrossRef]
21. Pandey, A., Shettigar, J., & Bose, A. (2021). Evaluation of the inflation Forecasting Process of the Reserve Bank of India: A Text Analysis approach. *SAGE Open*, *11*(3). [CrossRef]
22. Pollock, D. S. G. & University of Leicester. (2008). Statistical Fourier Analysis: Clarifications and Interpretations. In *Working Paper* (No. 08/36). [CrossRef]
23. Shibata, I. (2019, December 20). *Labor Market Dynamics: A hidden Markov approach*. IMF. [CrossRef]
24. Thalassinakis, E. (2025). An In-Depth investigation of the Riemann Zeta Function using infinite numbers. *Mathematics*, *13*(9), 1483. [CrossRef]
25. Vincent, P. (2023). A Comparative Study on Suitability of AHP and TOPSIS for Identifying optimal Conceptual Design of Bearing Puller. *Journal of Engineering Research and Reports*, *25*(12), 194–205. [CrossRef]
26. Yuan, Y., & Mitra, G. (2016). Market regime identification using hidden Markov models. *SSRN Electronic Journal*. [CrossRef]
27. Zhang, Y., Lo, A. W., & Terman, C. J. (2016). Spectral Analysis of High-Frequency Finance. In Massachusetts Institute of Technology, *Department of Electrical Engineering and Computer Science*. [CrossRef]

28. Zhu, S., Li, D., Feng, H., Gu, T., & Zhu, J. (2019). AHP-TOPSIS-Based Evaluation of the relative performance of multiple neighborhood renewal projects: a case study in Nanjing, China. *Sustainability*, *11*(17), 4545. [CrossRef]
29. Cochrane, J. H. (1997). *Time series for macroeconomics and finance* (Spring 1997; Pictures added Jan 2005). University of Chicago. [CrossRef]
30. Granger, C. W. J., & Watson, M. W. (1984). Time series and spectral methods in econometrics. In Z. Griliches & M. D. Intriligator (Eds.). *Handbook of Econometrics* (Vol. 2, pp. 980–1019). Elsevier. [CrossRef]
31. Rahman, M. (2024). Comparative analysis of AHP and TOPSIS methods in retail business location selection decision support system. *Journal of Engineering and Computational Economics, 2*(2). [CrossRef]
32. National Bank of Kazakhstan. (2024). *Selective-Combined Inflation Forecasting System (SSCIF)*. [CrossRef]
33. Ahmad, S., Ahmed, W., Choudhri, E. U., Pasha, F., & Tahir, A. (2017). *A model for forecasting and policy analysis in Pakistan: The role of government and external sectors*. International Growth Centre. [CrossRef]
34. Al-Mashat, R. A. (2018). Nuts and Bolts of a Forecasting and Policy Analysis System. In *Advancing the Frontiers of Monetary Policy*. USA: International Monetary Fund. [CrossRef]
35. International Monetary Fund. (2024). *Ghana: Technical assistance report on developing the forecasting and policy analysis system (FPAS) at the Bank of Ghana* (Technical Assistance Report No. 2024/080). [CrossRef]
36. Laxton, D., Rose, D., & Scott, A. (2009). *Developing a structured forecasting and policy analysis system to support inflation-forecast targeting (IFT)* (IMF Working Paper No. 09/65). [CrossRef]
37. National Bank of Georgia. (n.d.). *Macroeconomic statistics*. [CrossRef]
38. National Statistics Office of Georgia (Geostat). (n.d.). *Macroeconomic indicators*. [CrossRef]